# Valence Band Onset and Valence Plasmons of $SnO_2$ and $In_{2-x}Sn_x O_3$ Thin Films

Shailendra Kumar[1], C. Mukherjee[2] and D. M. Phase[3]

[1]Indus Synchrotron Utilization Division, [2]Mechanical and Optical Support Section,

Raja Ramanna Centre for Advanced Technology, Indore-452013, India

[3]Inter University Consortium, DAVV Campus, Indore-452001, India

## Abstract

Valence band onset ($E_v$), valence band tail (VBT) and valence plasmons (VPs) have been studied as a function of sputtering of $SnO_2$ and $In_{2-x}Sn_xO_3$ (ITO) thin films, using ultraviolet photoemission spectroscopy (UPS). Decrease in $E_v$ with respect to the Fermi level and increase in the density of energy levels of VBT have been observed after 5 minutes of sputtering using $Ar^+$ ions (500V). Bulk and surface components of VPs of Sn, SnO and $SnO_2$ in sputtered $SnO_2$ thin films have been observed in UPS spectra. Similarly, bulk and surface components of VPs of In, Sn and oxygen deficient ITO in sputtered ITO thin films have been observed in UPS spectra. Possible roles of $E_v$ and increase in the density of energy levels of VBT are discussed in the mechanisms of current transport through heterojunctions of $SnO_2$ with semiconductors.

Highlights :

1. Valence bulk and surface plasmons of $SnO_2$ and $In_{2-x}Sn_xO_3$.
2. Localized surface plasmons of Sn and In nano-clusters.
3. Increase in the density of energy levels in valence band tail after sputtering.
4. Current transport through the heterojunction of TCO and semiconductor.

Corresponding author : shail@rrcat.gov.in (Dr Shailendra Kumar).

1. Introduction

Transparent conducting oxides (TCOs) are used in many photonic devices and large area solar cells. Among many TCOs, doped $SnO_2$ and $In_{2-x}Sn_xO_3$ (ITO) thin films are commonly used in solar cells. Surprisingly, the mechanism of current transport through the heterojunctions of TCO and semiconductors is an active topic of research even now. Many groups are working on the role of semiconducting layer of TCO/semiconductor heterojunctions in current transport [1-6].

Klein et al [7] reported variation in the work function and difference between the Fermi level ($E_F$) and top of the valence band ($E_v$) as a function of deposition parameters of TCO thin films, deposited by magnetron sputtering. Schade et al [8] studied changes in F doped $SnO_2$ (FTO) thin films after interaction with hydrogen plasma using transmission, X-ray photoelectron spectroscopy (XPS) and Auger electron spectroscopy (AES). They observed decrease in transparency and formation of Sn and SnO at the surface after interaction with hydrogen plasma. Detailed studies of the interface of $SnO_2$ with amorphous silicon hydrogen alloy (a-Si:H) and amorphous silicon carbon hydrogen alloy (a-SiC:H) had been done by various groups [9,10]. Properties of $SnO_2$/semiconductor interface depend on the sequence of deposition. For example, if amorphous a-Si:H is deposited by plasma enhanced chemical vapor deposition (PECVD) on $SnO_2$, then plasma will take away some oxygen from $SnO_2$ surface and clusters of SnO and Sn are formed. The presence of clusters of SnO and Sn will change the current transport mechanism through the interface. Further, if $SnO_2$ is deposited on a-Si:H layer, there is finite probability of breaking of SiH bond and formation of $SiO_x$, SnO and Sn at the interface of a-Si:H and $SnO_2$.

Batzill et al [11] did detailed study of surface electronic structure of $SnO_2$ after sputtering and annealing in different conditions. Review article by Batzill and Debold describes physical properties of SnO$_2$ crystals, thin films and nano structures [12]. Köver et.al. [13] studied E$_v$, electronic structures and VPs energies of Sn, polycrystalline SnO and SnO$_2$, and (110) natural single crystal of SnO$_2$ using XPS, AES and electron energy loss spectroscopy (EELS). Many groups also studied electronic structures, free electrons plasmon (FP) and VP energies of Sn, SnO and SnO$_2$ and of ITO by AES and EELS [14-19]. It is a topic of research to find out that, whether, the presence of nano clusters of SnO and Sn at the interface of SnO$_2$/semiconductor is useful for a device or not. These arguments hold also for other reduced TCO/semiconductor interfaces also.

The presence of nano-clusters of SnO and Sn, at the surface of SnO$_2$ thin film, will change the optical absorption and current transport mechanism through the interface. The energy of free electrons bulk plasmons (FBP) is about 0.5eV, for density of free electrons about 5x10$^{20}$ cm$^{-3}$ in the conduction band of SnO$_2$ and the energy of valence bulk plasmons (VBP, due to valence electrons participating in the bonding) is about 18eV [11-14]. The experimentally observed energy of VBP in Sn is about 13eV and theoretically calculated values are about 11.4eV for bcc Sn and about 9eV for β-Sn [13,19, 20 ]. Energies of localized surface plasmons (LSP) of nano clusters will depend upon sizes and shapes of nano-clusters and dielectric properties of the surrounding materials. The LSP resonance frequency ($\omega_{lsp}$), for spherical nano particles, embedded in infinite, uniform host material, is approximately given by [21]

$$\omega_{lsp} = \frac{\omega_b}{\sqrt{\varepsilon_{inter}+2\varepsilon_{host}}} \qquad (1).$$

Here, $\omega_b$ is bulk plasmon frequency, $\varepsilon_{inter}$ is the constant interband screening in the metal and $\varepsilon_{host}$ is the dielectric constant of the surrounding medium. Kjeldsen et al [21] measured LSP of metallic Sn nano-clusters embedded in amorphous Si and SiO$_2$. They reported the LSP resonance of Sn nano-clusters around 5.5eV in $SiO_2$ and around 2.5eV in amorphous Si, measured by optical spectroscopy.

Prelim results related with plasmon peaks due to $SnO_2$, SnO and Sn at the surface of F doped $SnO_2$ (prepared by sputtering) and changes in $E_v$ as a function of sputtering time had been published in a conference proceedings [22]. In the present work, $SnO_2$ thin films prepared by pulsed laser deposition (PLD), commercial F doped $SnO_2$ thin films and ITO thin films prepared by sputtering were chosen for studies using ultraviolet photoelectron spectroscopy (UPS). A comparative study of un-sputtered and 5m sputtered surfaces has been done, using UPS spectra. No annealing was done after sputtering. Aim is to investigate the effect of low power sputtering of about 5 m on $E_v$ and valence band tails (VBT). Results related with $E_v$, VBT, VPs and LSP of Sn, $SnO_x$ and $In_{2-x}Sn_xO_{3-y}$ nano-clusters are presented.

## 2. Experimental

$SnO_2$ thin films were grown by PLD. The ceramic target, used for $SnO_2$ film growth by PLD, was prepared by the standard solid state reaction technique. A KrF excimer laser source was used to ablate the target. The substrate (quartz glass) was cleaned by the chemical method thoroughly and the chamber was evacuated at a base pressure of $2 \times 10^{-6}$ Torr before deposition. Depositions were performed at a substrate temperature of $700^{o}C$ and an oxygen partial pressure of $1 \times 10^{-4}$ Torr, while target to substrate distance was 5cm. The laser energy density and the pulse repetition rate were kept at 1.8 $Jcm^{-2}$ and 10 Hz, respectively. Samples were cooled in the same pressure as used during the deposition, at the rate of $20^{o}C$ $min^{-1}$. Thicknesses of the films were about 200nm as determined using a Talystep profilometer. These films are polycrystalline.

RF sputtering was used for the deposition of ITO films. Quartz and Si wafer were used as substrates. Substrate preparation and cleaning prior to deposition plays key role in the deposited film quality. The substrates were cleaned ultrasonically in alcohol/acetone and in de-ionized water before the deposition. To get uniform film, substrates were rotated at 5rpm during deposition. The film depositions were carried out in an automated sputtering system (Elerttorava spa., Italy). 99.99% pure ITO target (ACI Alloys Inc.) was used for deposition. The vacuum chamber was pumped down below $7\times10^{-7}$ mbar before the deposition. Substrate temperature was maintained at room temperature, during the deposition. High purity (99.9995%) $O_2$ (3sccm) and Ar (100 sccm) were used. Chamber pressure was maintained at 3.0 mTorr with the help of a baratron and throttle valve combination. The deposition rate was 0.1nm/s at 100W RF power (Huttinger), measured by quartz crystal monitor (Inficon, IC/5). Prior to deposition target conditioning was done by pre-sputtering for 5mins. Thickness of the film was 85nm. Optical transmission and reflectivity spectra were recorded in CARY5000 spectrophotometer. Films were annealed at $150^o$ C for 2Hrs. at ambient environment. Commercial F doped $SnO_2$ films were prepared by sputtering.

UPS spectra had been measured on angle integrated photoelectron spectroscopy (AIPES) beam line of INDUS-1 synchrotron source. Vacuum in the AIPES experimental chamber was better than $2\times10^{-9}$ Torr. Other details of AIPES experimental chambers are given in previously published paper [23]. $Ar^+$ ions sputtering was done at $2\times10^{-5}$ Torr partial pressure of argon, at room temperature, in the preparation chamber separated by the gate valve from the experimental chamber. Energy of $Ar^+$ ions was 500 eV and current was 5μA. After sputtering, samples were transferred to experimental chamber and UPS spectra were recorded within half an hour after sputtering. Photoelectron counts are collected in steps of 20 meV with pass energy of 20eV. Polycrystalline Au foil was also loaded along with $SnO_2$ samples. UPS spectrum from the Au foil is used as a reference to characterize the monochromatic synchrotron light and the Fermi energy level ($E_F$) of the metallic sample holder.

## 3. Results and Discussion

### 3.1 $SnO_2$ thin films

UPS spectra, after base line subtraction in the BE range 25-60 eV, are shown for PLD deposited un-sputtered and sputtered $SnO_2$ thin films in Fig. 1(a) and (b), respectively. The energy of incident photons is 100eV. Peak $P_1$ and broad peak $P_2$ at binding energy (BE) = 31.3 eV and 44 eV, respectively are due to loss of kinetic energy (KE) of Sn-4d core photoelectrons after scattering with plasmons. The difference between Sn-4d core level and peak $P_1$ is 4.7 eV. In literature, some groups have assigned this peak to inter-band Sn-5s to Sn-5p transitions [13]. However, the source of this peak may also be due to LSP of Sn nano clusters partially embedded within $SnO_2$ at the surface. The reported energy of the LSP of nano clusters of Sn embedded in $SiO_2$ is about 5.5eV [21]. The refractive index ($\eta$) of $SiO_2$ is 1.5, while $\eta$ of $SnO_2$ is about 2.0, hence, the energy of LSP of Sn nano clusters embedded in $SnO_2$ will be little less than 5eV. In published literature, related with UPS and EELS spectra of $SnO_2$, some groups have reported the presence of the peak $P_1$ and others have not reported it [10-19]. If it is due to the inter-band transition, then it should have been present in all samples. Therefore, the peak $P_1$ is due to scattering of photoelectrons with LSP of Sn nano clusters embedded in $SnO_2$. Broad peak $P_2$ is due to combined effect of bulk and surface components of VPs at 19eV and 13.4eV, respectively of $SnO_2$.

The shape of the plasmon peaks of sputtered $SnO_2$ thin film, Fig. 1(b), has changed as compared to un-sputtered surface. Two new maxima have appeared and these are indicated by peaks $P_3$ and $P_4$ in Fig. 1(b). The separation of $P_3$ and $P_4$ peaks from the Sn-4d core level maximum are 10 and 14 eV, respectively. These values match with energies of VBPs of SnO and Sn, respectively. Further, the energies of valence surface plasmons (VSPs = VBPs/1.41) of $SnO_2$, SnO and Sn are 12.8, 8.3 and 10.2 eV, respectively. The broad plasmons peak is sum of VBPs and VSPs of $SnO_2$, SnO and Sn. The relative intensity of VBPs due to $SnO_2$ at $P_2$ has reduced for sputtered surface. Hence, these results show that nano-clusters of SnO and Sn are present on the sputtered surface of $SnO_2$.

Fig. 2 (a) and (b) show expanded valence band onsets (of figure 1) of un-sputtered and 5 minutes sputtered surfaces, respectively of PLD deposited $SnO_2$ thin film. The intersection point of tangents to the leading edge of the valence band onset and the zero line is taken as the valence band onset point. The difference between the Fermi level (BE=0) and the intersection point gives position of the valence band maximum with respect to the Fermi level ($E_F - E_v$) [24]. The magnitude of ($E_F - E_v$) for un-sputtered surface of $SnO_2$ is 3.7±0.05 eV. Error in the magnitude of ($E_F - E_v$) is not written in rest of the manuscript. The optical band gap of PLD deposited $SnO_2$ thin film is also 3.7 eV within experimental errors. This shows that the Fermi level is at the bottom of the conduction band in PLD deposited $SnO_2$ thin films. It is to be noticed that the measurable band tail is up to 2.0 eV for un-sputtered surface, Fig. 2(a). The shape of the valence band onset has drastically changed for 5 minutes sputtered surface as shown in Fig. 2(b). On the sputtered surface, there are two distinct valence band onsets. Two tangents to valence bands onsets are intersecting the zero line at 1.8 and 3.2 eV, respectively, indicating presence of two phases. Measurable VBT are up to 1eV. These changes in $E_v$ of sputtered surface are important from device point of view. These onsets are due to presence of oxygen deficient phases of $SnO_x$ and Sn at the surface.

Effect of sputtering on the valence band onsets for a commercial thin film of F doped $SnO_2$ are shown in Fig. 3 for comparison with PLD deposited $SnO_2$ thin films. In Fig. 3, curve 1 is for un-sputtered thin film and curve 2 is for 5 minutes sputtered thin film. Details of effects of sputtering on plasmons peaks for $SnO_2(F)$ thin films had been published earlier [22]. Values of ($E_F$ - $E_v$) for un-sputtered surface and 5 minutes sputtered surface are 3.2 eV and 2.7 eV, respectively. VBT are up to 1.0 eV below $E_F$ and the density of energy levels in the valence band tail is larger for the sputtered surface in comparison to the un-sputtered surface. These observations show that the valence band onset results for PLD deposited $SnO_2$ and F doped $SnO_2$ thin films depend upon the deposition process and sputter cleaning. Sputtered cleaned surface of $SnO_2$ have nano clusters of SnO and Sn and their concentrations determine ($E_F$ - $E_v$) and VBT. These observations are important from the point of view of the role of valence band tails in the current flow mechanisms through the heterojunction of $SnO_2$/semiconductors.

## 3.2 ITO thin films

UPS spectra in Fig. 4(a) and (b) of an ITO thin film show In-4d, Sn-4d core levels and plasmons peaks of un-sputtered and 5 minutes sputtered surfaces, respectively. The energy of incident photons is 100eV. Reported experimentally measured energy of bulk VPs of ITO, by EELS, is 19 eV [19]. Surprisingly, this value is similar to reported energy of bulk VPs of $SnO_2$ thin films [14-17]. Expected peaks due to bulk and surface components of VPs of ITO, with respect to In-4d core level, should be at BE=38 eV and 32 eV, respectively. Similarly, expected peaks due to bulk and surface components of VPs, with respect to Sn-4d core level, should be at BE= 46 eV and 40 eV, respectively. The broad peak (22-50 eV) in UPS spectra of un-sputtered surface is combination of bulk and surface components of VPs of ITO with respect to both In-4d and Sn-4d core peaks. Small hump near BE = 46 eV in Fig. 4(b) is due to bulk component of VPs, due to scattering of Sn-4d photoelectrons with VPs. In Fig. 4(b), small hump near BE= 23.5 eV is probably due to LSP of In nano clusters partially embedded in ITO at the surface. The peak due to LSP of Sn nano clusters, expected at BE=31.5 eV, is merged in the broad peak. The maxima of the broad plasmon peak of sputtered surface, Fig. 4(b), is shifted towards the lower BE in comparison to un-sputtered surface, Fig. 4(a), and this shows the oxygen deficiency at the sputtered surface of ITO.

Fig. 5 shows expanded region of the valence band onsets of Fig.4. ($E_F - E_v$) is 2.8 eV and measurable VBTs are up to 1eV for un-sputtered surface. The valence band of the sputtered surface has two distinct parts, different from the un-sputtered surface. The intersection points of tangents of valence band onsets with zero line are at 0.8eV and 2.75eV in Fig. 5(b). The absorption spectrum of ITO film (thickness = 85 nm) was also measured in the 1eV-5eV range. The band gap of ITO thin film is 3.15 eV as shown from the absorption spectrum in Fig.6. There is finite measurable absorption in the sub band gap region of ITO film, supporting the observation of VBTs in UPS spectrum. Next, role of VBTs in F doped $SnO_2$, in current transport through heterojunction of $SnO_2$/a-Si:H(p) is discussed as a representative case and these arguments will hold good for heterojunctions of $SnO_2$ and ITO with other p doped semiconductors.

3.3 Current transport through heterojunction

Role of conduction and valence bands tails in the current transport through the heterojunction of $SnO_2$ and a-Si:H is discussed here, as these heterojunctions are important in solar cells and other devices. A possible schematic of a solar cell is shown in Fig. 7 with $SnO_2$/a-Si:H(p) and a-Si:H(n)/ $SnO_2$ junctions. The flow of holes from a-Si:H(p) to $SnO_2$ and flow of electrons from a-Si:H(n) to $SnO_2$ will determine the magnitudes of the short circuit current ($I_{sc}$) and open circuit voltage ($V_{oc}$), in the presence of light. In the case of $SnO_2(F)$, the valence band tail is observed up to 1eV below the $E_F$ for un-sputtered as well as sputtered surfaces. For un-sputtered surface of $SnO_2(F)$, measured value of ($E_F - E_v$) is 3.2eV ( Fig. 3) and the optical band gap of $SnO_2(F)$ is 3.6 eV. $E_F$ in the neutral bulk of $SnO_2(F)$, away from the space charge region of the interface, is about 0.4eV above the conduction band minimum ($E_c$) for free electrons density of about $5 \times 10^{20}$ cm$^{-3}$ [25, 26]. Hence, there is upward band bending at the surface of $SnO_2$, considering that the optical gap ($E_c$-$E_v$) is 3.6eV and ($E_F - E_v$) is 3.2eV at the surface.

There is downward band bending at the surface of a-Si:H(p) and upward band bending at the surface of a-Si:H(n) [27]. A possible bands alignment at the heterojunction of $SnO_2$/a-Si:H(p) is shown in Fig. 8(a). There is a potential barrier of about 2.8 eV for flow of holes from a-Si:H(p) to $SnO_2$. Now, the query is about the roles of potential barrier at the $SnO_2$/a-Si:H(p) and the presence of VBTs in flow of holes from a-Si:H(p) to $SnO_2$. It is known from published literature that $SnO_2$ is oxygen deficient at the interface of $SnO_2$/a-Si:H. Hence, UPS spectra of sputtered $SnO_2$ surface can be utilized for understanding the role of VBTs in the mechanism of the current flow through the interface of $SnO_2$/a-Si:H. It is clear from the UPS spectra, of the sputtered $SnO_2(F)$ surface, Fig. 3, that there is measurable continuous distribution of VBTs above $E_v$, up to 1.0 eV below $E_F$. The presence of continuous VBTs about 1.0 eV below $E_F$ in the sub band gap region will give rise to tunneling and hopping conduction as a dominant mode of holes transport across the interface. Holes will tunnel from a-Si:H(p) to VBTs in $SnO_2(F)$, followed by recombination with electrons from the conduction band of $SnO_2(F)$, as shown in Fig. 8(a). The magnitude of the tunneling current will be determined by the density of the occupied energy levels in VBTs of $SnO_2(F)$. It is clear from the measured UPS spectra of $SnO_2$ and ITO that sputtering of only 5 minutes increases the density of VBTs. Hence, the sputtered surfaces of $SnO_2$ and ITO, at the interface, may be more useful for flow of holes from a-Si:H(p) to $SnO_2$. In the case of ITO film, it is clear from Fig. 5 and Fig. 6 that finite distribution of filled energy levels is present up to 0.8eV below $E_F$ at the surface. Hence, above discussion holds good also for ITO/a-Si:H(p) interface.

The flow of electrons from a-Si:H(n) to $SnO_2$(F) at the a-Si:H/ $SnO_2$(F) interface will be governed by the barrier height, as shown in Fig. 8(b). Flow of electrons through the interface of a-Si:H(n) and $SnO_2$(F) is by tunneling. As electrons are tunneling from tails below $E_c$ of a-Si:H(n) to unoccupied energy levels within the conduction band of $SnO_2$(F), the current will be determined by the width of the barrier. The change in the density of states within the conduction band of $SnO_2$(F), after sputtering will not be much. The width of the barrier is less for un-sputtered surface in comparison to sputtered surface of $SnO_2$(F). Hence, un-sputtered surface of $SnO_2$(F) will be more useful for the flow of electrons from a-Si:H(n) to $SnO_2$(F). Next, plasmonic role of nano-clusters of Sn and In sandwiched at the interface of TCO/a-Si:H is discussed.

The approximate energy of LSP is given by equation (1) and reported value of LSP for Sn nano-clusters in a-Si is 2.5eV [21]. The expected value of LSP for Sn and In nano clusters in $SnO_2$ will be also in the range 2-3eV, as experimentally measured bulk plasmon energies of Sn and In are in the range 9 -13 eV. Hence, LSP energies of Sn and In nano clusters, embedded in semiconductors, can be varied in the visible energy range. The utilization of plasmonic properties of Sn and In nano clusters, in place of noble metals Au and Ag nano clusters, will be cheaper. Nano clusters of Sn and In can be formed by a few minutes of sputtering of ITO or $SnO_2$ thin films. This can be achieved without deposition of noble metals in the fabrication of photovoltaics. The LSP energies of nano clusters of $SnO_2$ and $In_{2-x}Sn_xO_3$, due to free electron bulk plasmon energies (about 0.5eV), will be about 0.1 eV in a-Si medium. This is far less than the useful spectrum range of Sun-light for photovoltaics. On the other hand, the LSP energies of nano clusters of $SnO_2$ and $In_{2-x}Sn_xO_3$ due to valence bulk plasmon energies (about 19 eV) will be about 4 eV in a-Si medium. This is more than the useful spectrum range of Sun-light for photovoltaics and photons of this energy will be absorbed in $SnO_2$ (ITO) thin film itself.

## 4. Conclusions

In conclusion, few minutes of sputtering of $SnO_2$ thin films decreases the magnitude of ($E_F$-$E_v$) and measurable valence band tails are present up to 1eV with respect to the Fermi level. In the case of ITO thin films, few minutes of sputtering shifts the valence band onset to 0.8 eV with respect to the Fermi level. Decrease in ($E_F$-$E_v$) and increase in the density of energy levels in valence band tails of sputtered $SnO_2$ (and ITO) will be useful for holes transport from a-Si:H(p) to $SnO_2$ (ITO). Few minutes of sputtering also creates nano-clusters of Sn, In and oxygen deficient $SnO_x$ and $ITO_x$. LSP energies of Sn and In nano-clusters, embedded in semiconductors (of optical band gaps about 1.5 - 1.8eV) are in the visible range of the solar spectrum and it is proposed that these can be utilized in increasing the efficiency of solar cells.

**Acknowledgment**

Authors are thankful to Mr K. Rajiv and Mr A. Wadikar for their help in experiments, and to Dr G. Lodha for his support in this activity.

**Figure captions**

Fig. 1 : UPS spectra of (a) un-sputtered and (b) 5 minutes sputtered surface of $SnO_2$ thin film deposited by PLD as a function of binding energy (B.E.). Along with Sn-4d core peak, peaks $P_1$, $P_2$, $P_3$ and $P_4$ are due to confined localised plasmons of Sn nano clusters, combined bulk and surface valence plasmons of $SnO_2$, valence plasmons of SnO and valence plasmons of Sn, respectively. More details are in text.

Fig.2 : UPS spectra showing valence band onset and valence band tails of (a) un-sputtered and (b) sputtered of $SnO_2$ thin film deposited by PLD. $E_F$ is the Fermi level, $E_v$ is the valence band onset of un-sputtered surface, $E_{v1}$ and $E_{v2}$ are valence band onsets of two phases present on sputtered surface, $E_t$ is the onset of valence band tails. Superscript u and s are for un-sputtered and sputtered surfaces, respectively.

Fig.3 : Valence band onset and valence band tails of UPS spectra of (a) un-sputtered and (b) 5 minutes sputtered commercial F doped $SnO_2$ thin film. Symbols have same meaning as in Fig.2.

Fig.4 : UPS spectra of (a) un-sputtered and (b) 5 minutes sputtered surface of ITO thin film deposited by RF sputtering. In-4d, Sn-4d core peaks along with broad plasmon peak are shown.

Fig.5 : Expanded region of valence band onset and valence band tails of UPS spectra of un-sputtered and sputtered of ITO thin film deposited by RF sputtering. Symbols have same meaning as in Fig.2.

Fig.6 : Absorption spectrum of ITO thin film measured by transmission. Optical band-gap ($E_g$) is 3.15eV and absorption from band tails is quite large.

Fig.7 : Schematic of an amorphous silicon based solar cell band diagram showing TCO/a-Si:H(p) and a-Si:H(n)/TCO junctions. $E_c$, $E_v$ and $E_F$ are bottom of the conduction band, top of the valence band and the Fermi level, respectively, in dark.

Fig.8 : Schematic bands alignment at (a) $SnO_2(F)$/a-Si:H(p) interface and (b) $SnO_2(F)$/a-SiH(n) interface. At $SnO_2(F)$/a-Si:H(p) interface, excess holes from a-Si:H(p) tunnel to valence band tail of $SnO_2(F)$ followed by recombination of electrons from the conduction band with holes in valence band tail in $SnO_2(F)$. At $SnO_2(F)$/a-Si:H(n) interface, excess electrons from the conduction band tail of a-Si:H(n) tunnel to the conduction band of $SnO_2(F)$.

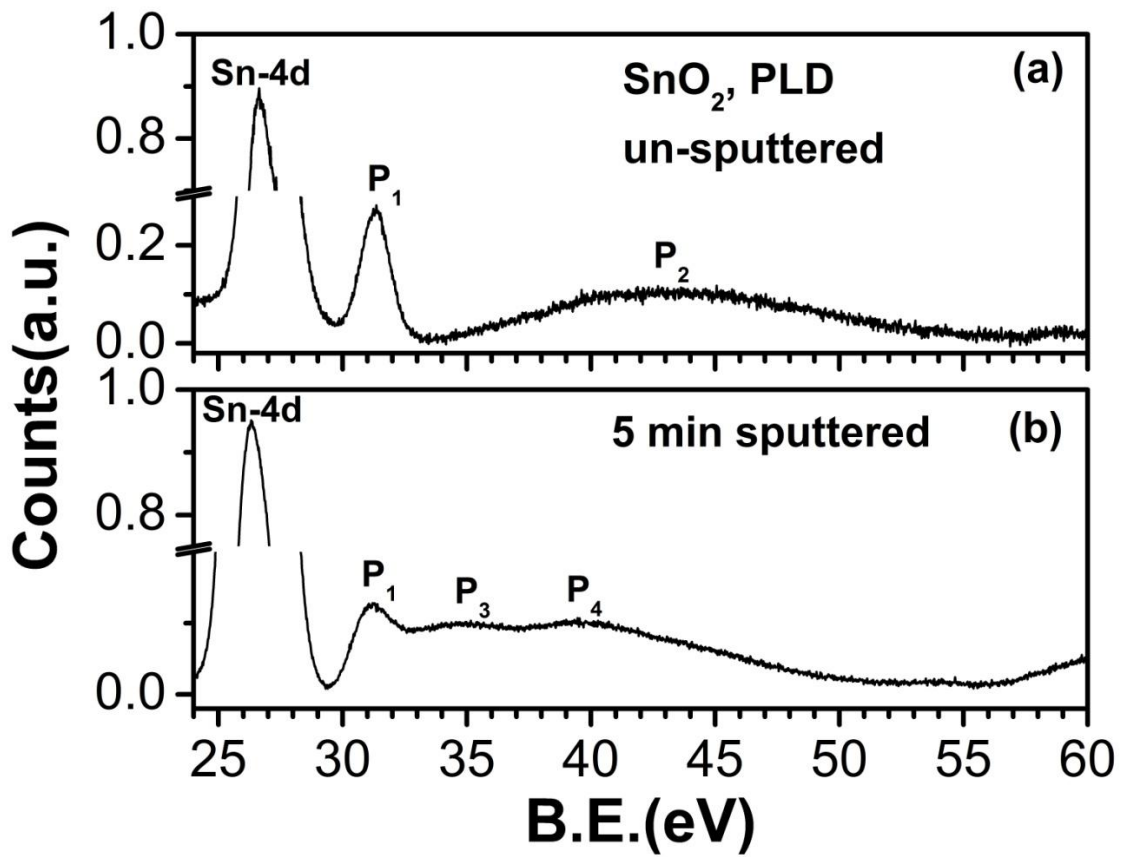

Fig. 1

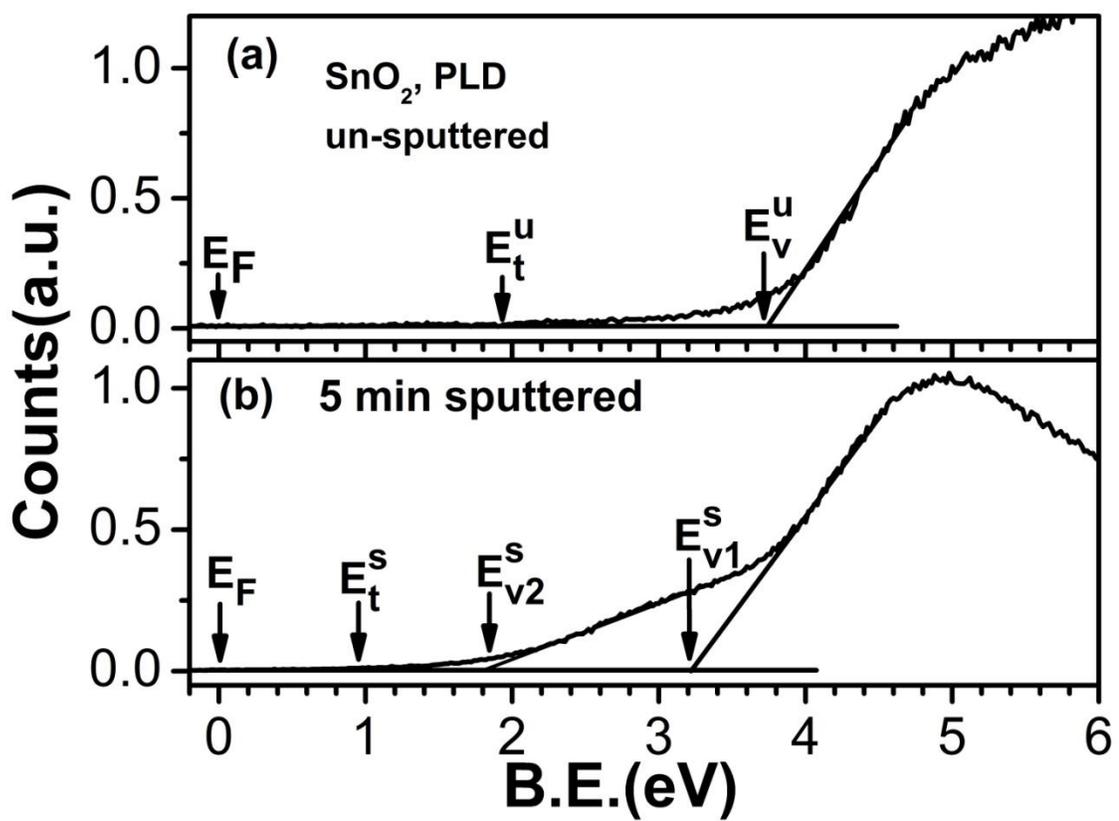

Fig. 2

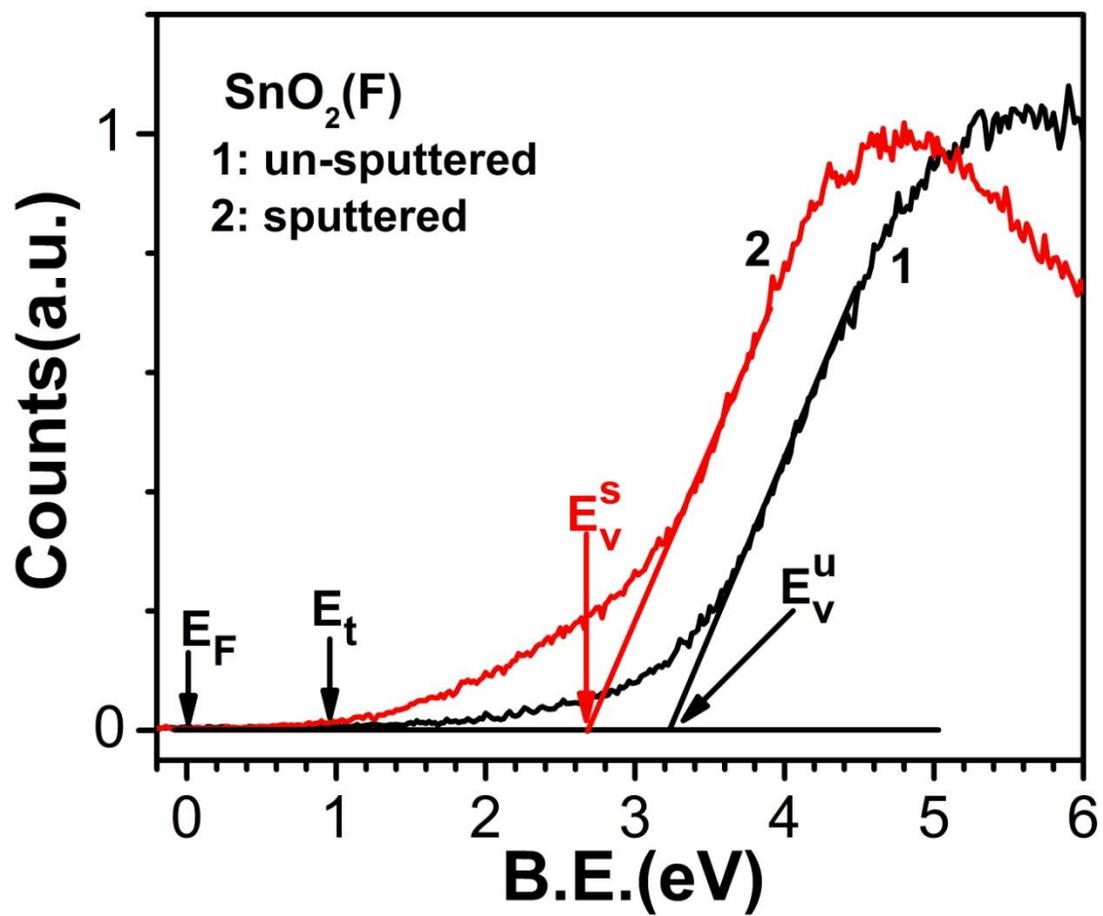

Fig. 3

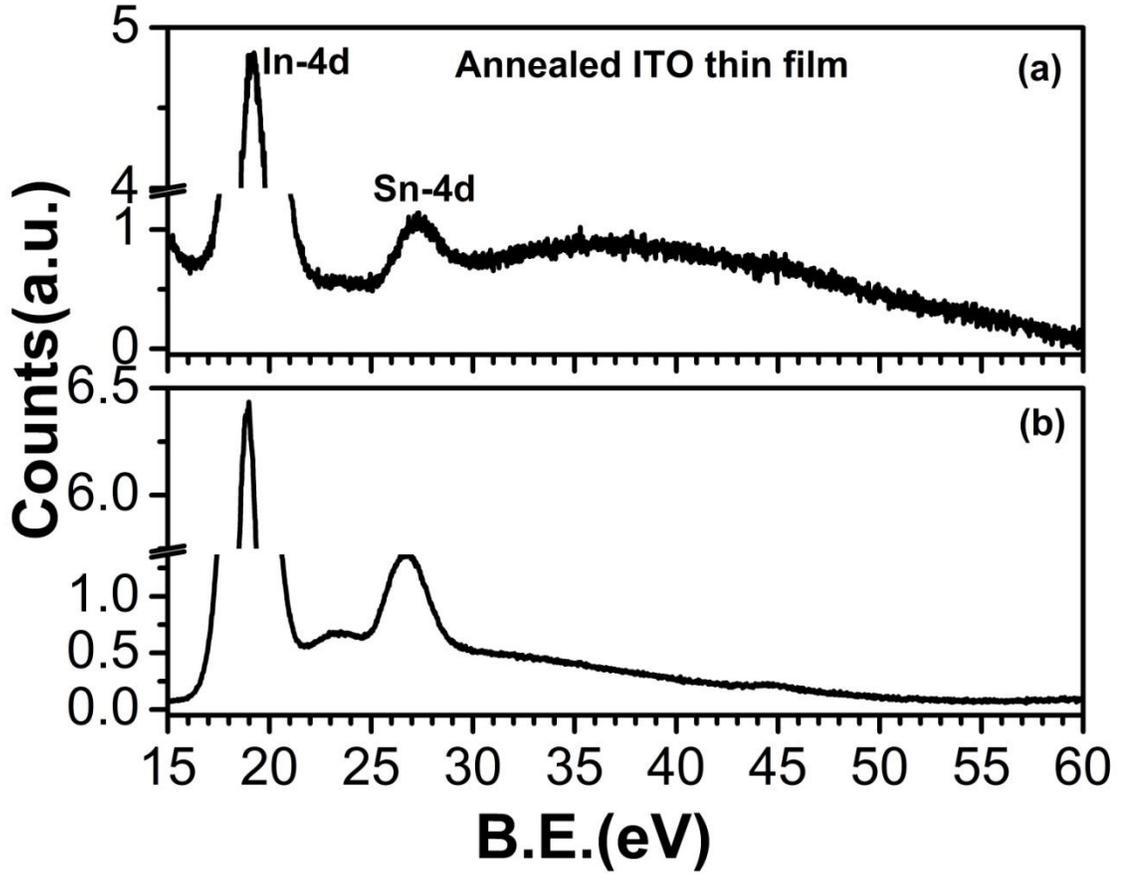

Fig. 4

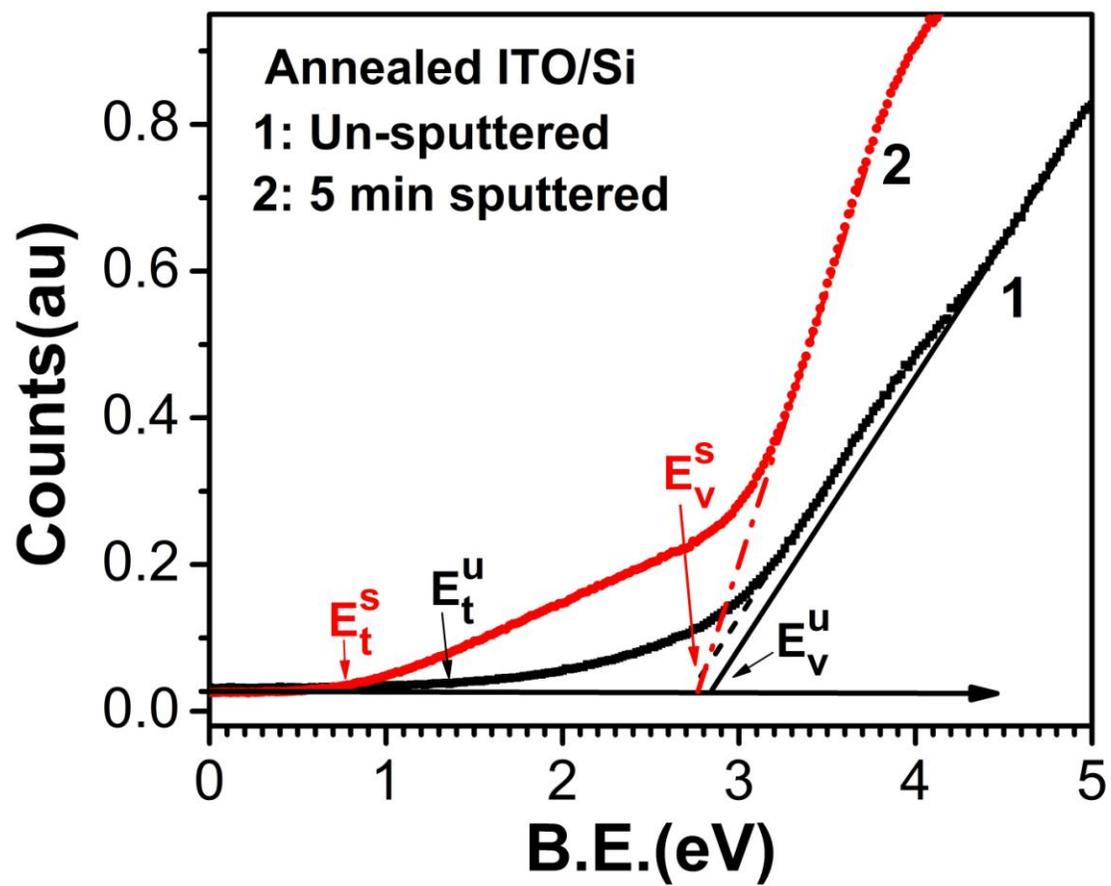

Fig. 5

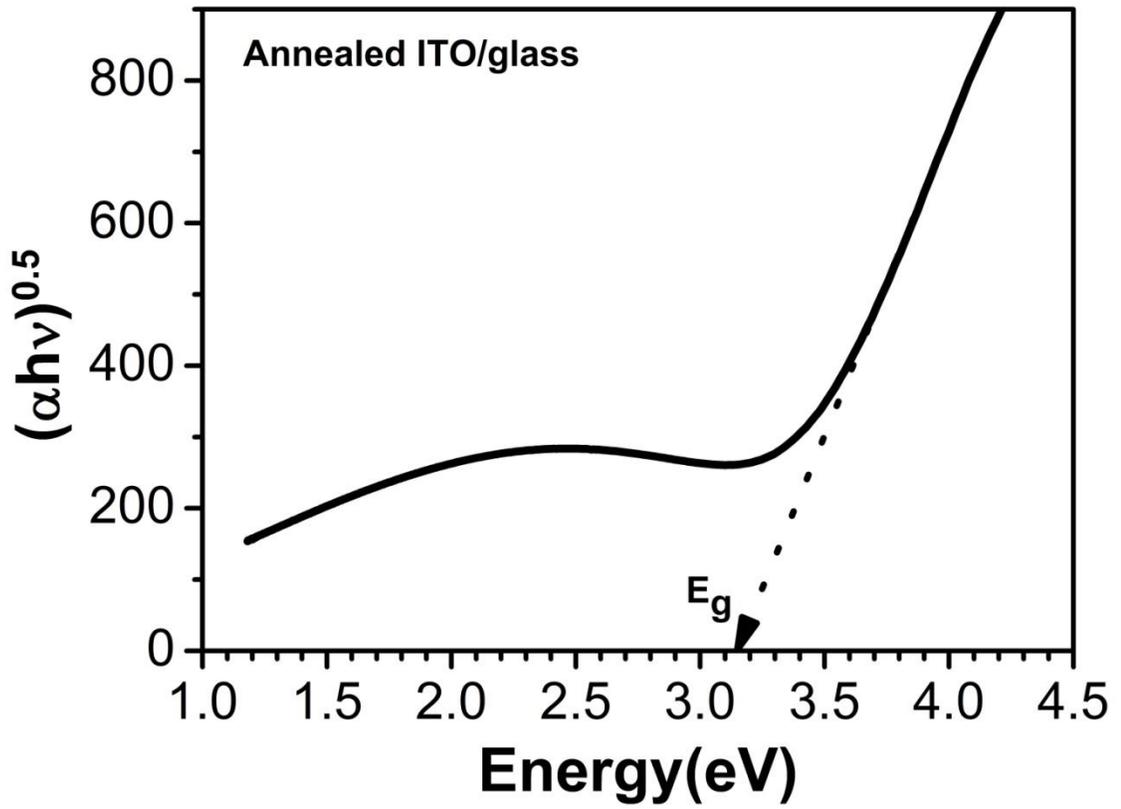

Fig. 6

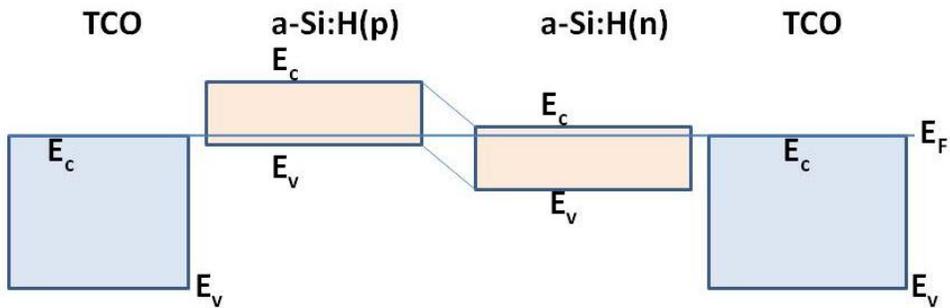

Fig. 7

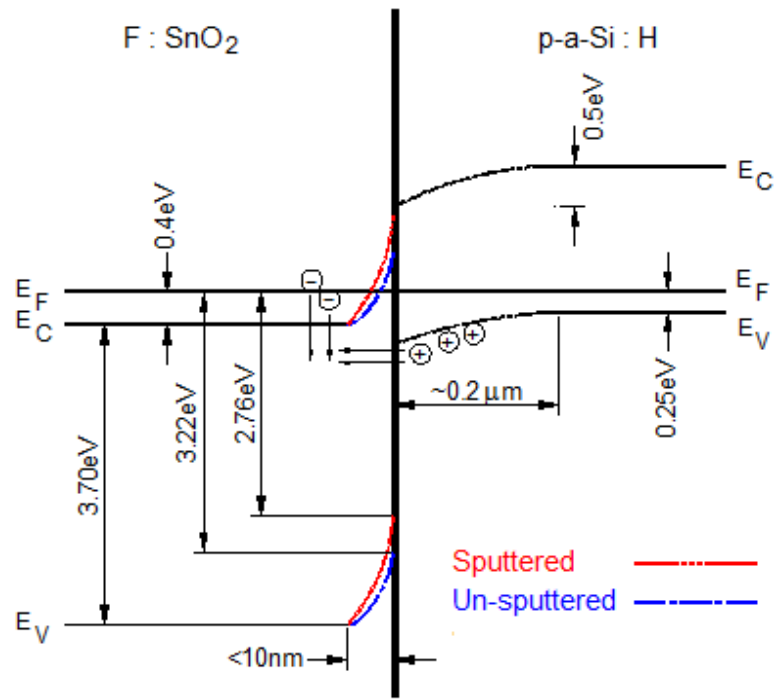

Fig. 8 (a)

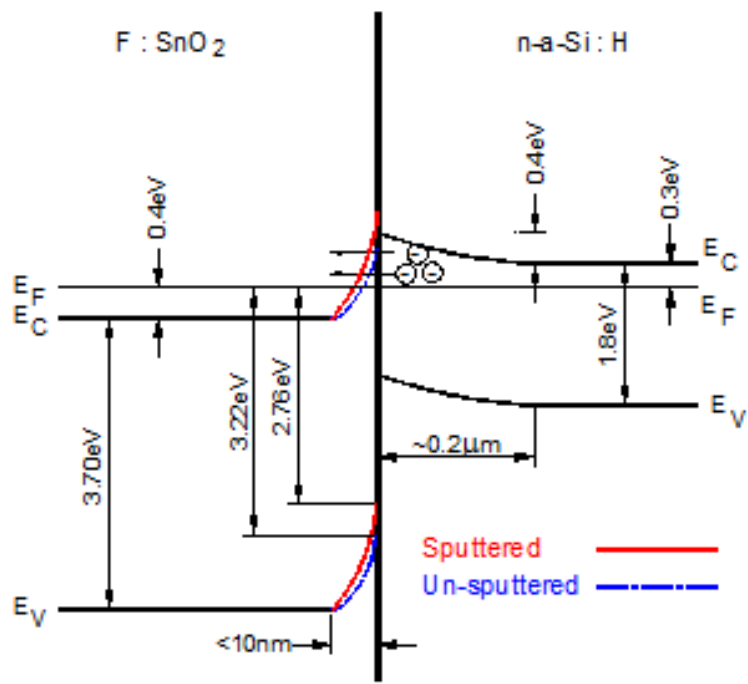

Fig. 8 (b)